\documentclass[12pt]{article}
\usepackage{geometry}
\usepackage{a4}
\usepackage{graphicx}
\usepackage{epsf}
\usepackage{amsmath}
\usepackage{amssymb}
\usepackage{cite}
\newcommand{\be}{\begin{equation}}
\newcommand{\ee}{\end{equation}}

\newcommand{\Rmnum}[1]{\expandafter\@slowromancap\romannumeral #1@}
\newcommand{\bea}{\begin{eqnarray}}
\newcommand{\eea}{\end{eqnarray}}

\begin{document}
\def\A{{\mathbb{A}}}
\def\B{{\mathbb{B}}}
\def\C{{\mathbb{C}}}
\def\R{{\mathbb{R}}}
\def\s{{\mathbb{S}}}
\def\T{{\mathbb{T}}}
\def\Z{{\mathbb{Z}}}
\def\W{{\mathbb{W}}}
\begin{titlepage}
\title{Information Geometry, Phase Transitions, and Widom Lines : Magnetic and Liquid Systems}
\author{}
\date{
Anshuman Dey, Pratim Roy, Tapobrata Sarkar
\thanks{\noindent E-mail:~ deyanshu, proy, tapo @iitk.ac.in}
\vskip0.4cm
{\sl Department of Physics, \\
Indian Institute of Technology,\\
Kanpur 208016, \\
India}}
\maketitle
\abstract{
\noindent
We study information geometry of the thermodynamics of first and second order phase transitions, and beyond criticality, in magnetic and 
liquid systems. We establish a universal microscopic characterization of such phase transitions via the equality of correlation lengths $\xi$ in coexisting phases,
where $\xi$ is related to the scalar curvature of the equilibrium thermodynamic state space.  
The 1-D Ising model, and the mean-field Curie-Weiss model are discussed, and we show that information geometry correctly describes the phase
behavior for the latter. The Widom lines for these systems are also established. We further study a simple model for the thermodynamics of liquid-liquid phase 
co-existence, and show that our method provides a simple and direct way to obtain its phase behavior and the locations of the Widom lines. Our analysis points 
towards multiple Widom lines in liquid systems.}
\end{titlepage}

\section{Introduction}

The physics of phase transitions \cite{callen} has been a fascinating area of research for more than a century, starting from the celebrated van der Waals
equation for liquid-gas systems. Application of information geometric methods to such studies are relatively recent, but have yielded several useful
insights especially in the context of liquid-gas co-existence \cite{brodyhook}. This method involves the underlying Riemannian geometry of the 
equilibrium thermodynamic state space,
and was mainly initiated through the work of Weinhold \cite{weinhold} and Ruppeiner \cite{rupp}.\footnote{Information geometry in the context of thermodynamics
is also called ``thermodynamic geometry.''}
The key idea here is to utilise the positivity condition on the Hessian of
the entropy for a thermodynamic system in equilibrium, so as to define a Riemannian metric on the thermodynamic state space. In a striking conjecture, 
Ruppeiner proposed, via the theory of Gaussian fluctuations, that the scalar curvature $R$ (or, more appropriately, $|R|$), arising out of such a metric is related to the 
correlation length $\xi$ of the system, an
idea that has since been tested in a variety of models. Indeed, it has been established by now that the scalar curvature of the thermodynamic state space diverges at critical points, 
for a wide variety of systems that exhibit second order phase transitions.\footnote{Information geometry of quantum phase transitions has also been an area
of extensive research, see e.g. \cite{zanardi}.}

Most of the analysis done in the past dealt with the geometry of thermodynamics at or near criticality. However, it has
recently been established \cite{rsss} that methods of information geometry can be used to study discontinuous first order phase transitions, via the equality of the correlation lengths
of the coexisting phases, in liquid-gas systems. In \cite{rsss}, such equality has, in fact, been verified with experimental data from NIST \cite{nist}.
Apart from providing an alternative route to study phase transitions that bypasses several long-standing problems 
in standard thermodynamics, this approach has the potential of analytically predicting the ``Widom line'' \cite{stanley},\cite{stanley2}, a conjectured continuation of phase  co-existence
that extends beyond criticality, and distinguishes between supercritical ``phases'' of a system. The presence of the Widom line, which is
usually defined as the locus of maxima of the correlation length has recently been experimentally established \cite{simeoni}, and certainly puts information theoretic
studies of phase transitions in perspective. Note that very close to criticality, all response coefficients scale as powers of the correlation length, and hence the loci of
maxima of these serve as equivalent definitions of the Widom line \cite{stanley}. Slightly away from criticality, however, this is not the case and our method provides a way
of locating the Widom line as per its actual definition, even away from criticality. 

Whereas methods of information geometry in liquid-gas systems have been firmly established by now, magnetic and liquid-liquid systems have been much less studied. 
Although it was known that the thermodynamic scalar curvature can be calculated for some magnetic systems, 
and these show the standard divergences associated with a second order critical point, it is not known how
Riemannian geometry captures first order transitions in these systems. Even less is known about the information geometry of liquid systems and its role in predicting phase 
behavior therein. Here we address these issues and show that geometrical techniques can be effectively used to study phase behavior and the Widom line in these cases as well. 
Our results here, in conjunction with those for liquid-gas systems analyzed in \cite{rsss} firmly establish the usefulness and applicability of information geometry in 
mean field thermodynamic systems, on which experimental predictions can be based. 
We note here however, that although the equality of correlation lengths in co-exisiting phases in liquid-gas systems can be understood \cite{rsss} following the work of 
Widom \cite{widom}, such a picture has not be rigorously developed for magnetic and liquid systems. We will proceed by assuming this, and show that it exactly 
reproduces known features in mean-field magnetic models. With this understanding, we then analyse liquid-liquid systems in a simplified mean-field model. 
The usefulness of our method is that it provides an analytic tool for predicting phase transitions and the Widom line in difficult situations where one otherwise needs to resort to 
molecular dynamics simulations \cite{stanley2}. 

This paper is organised as follows. In the first section, we briefly review the information geometry of equilibrium thermodynamics, and then proceed 
to study the exact solution of the 1-D Ising model in this context. This section is mostly devoted to establishing the notations and conventions that we follow in 
the rest of the paper, but we present an interesting observation regarding the Widom line in exactly solvable systems. 
Next, we move on to study phase behavior of magnetic systems in the mean-field Curie-Weiss model, and finally analyse liquid systems in section 3. 
For both these cases, the prediction of phase co-existence and the Widom line is elaborated using geometric methods. 
We end the paper with our conclusions and some future prospects. 

\section{Information Geometry and The 1-D Ising Model}

We begin with a very brief overview of information geometry. It will be enough for us to consider this in the context of equilibrium thermodynamics, a formulation
due to Ruppeiner \cite{rupp}. More details can be found in the excellent review of Brody and Hook \cite{brodyhook}. The main idea is to consider the 
line element due to a positive definite Riemannian metric which is defined by the Hessian of the entropy per unit volume $s$,
\begin{equation}
dl^2 = g_{ij}da^ida^j, ~~~g_{ij} = -\frac{1}{k_B}\left(\frac{\partial^2 s}{\partial a^i \partial a^j}\right)
\label{entropyrep}
\end{equation}
where $a^1$ and $a^2$ denote the internal energy and the particle number per unit volume (for magnetic systems, one has to appropriately use thermodynamic
quantities per unit spin). As shown by Ruppeiner \cite{rupp}, this introduces the concept of a distance in the space of equilibrium thermodynamic states, i.e, a large distance
between two such equilibrium states is interpreted as a small probability that these are related by a thermal fluctuation. 

One can consider various other forms of the metric related to the one given in eq.(\ref{entropyrep}) by Legendre transforms. Indeed, a choice of 
the thermodynamic potential in appropriate coordinates might render the metric diagonal, and simplify algebraic calculations. For example, for single component
fluids and magnetic systems, a particularly simple diagonal form of the metric can be used \cite{rupp}
\begin{equation}
dl^2 = \frac{1}{k_B T}\left(\frac{\partial s}{\partial T}\right)_{\rho} dT^2+ \frac{1}{k_B T}\left(\frac{\partial \mu}{\partial \rho}\right)_Td\rho^2
\label{line}
\end{equation}
where $\mu = \left(\frac{\partial f}{\partial \rho}\right)_T$, $f$ being the Helmholtz free energy per unit
volume (per unit spin for magnetic systems), and $\rho$ is the inverse of the volume. For magnetic systems,  the magnetization per unit volume plays
the role of $\rho$.

For such diagonal metrics, the
scalar curvature of the manifold takes a well known simple form,
\begin{equation}
R = \frac{1}{\sqrt{g}}\left[\frac{\partial}{\partial T}\left(\frac{1}{\sqrt{g}} \frac{\partial g_{\rho\rho}}{\partial T}\right) + 
\frac{\partial}{\partial \rho}\left(\frac{1}{\sqrt{g}} \frac{\partial g_{TT}}{\partial \rho}\right)\right]
\end{equation}
where $g_{TT}$ and $g_{\rho\rho}$ are the coefficients multiplying $dT^2$ and $d\rho^2$ respectively, in eq.(\ref{line}), and $g = g_{TT}.g_{\rho\rho}$. In passing,
we note that for positivity of the line element, $g_{TT}$ and $g_{\rho\rho}$ should be positive definite in the regions of interest. 
Non-diagonal forms of the metric (as in eq.(\ref{entropyrep})) can also be used, and give equivalent results for $R$.

By augmenting this geometric picture with Gaussian fluctuation theory, 
Ruppeiner conjectured that such a scalar curvature can, in fact, be related to the correlation length of the system \cite{rupp}, $|R| \sim \xi^d$,
where $d$ is the system dimension. We also mention that the geometry described here becomes trivial in the presence of a single fluctuating variable,
and meaningful results can only be obtained for two or more fluctuating thermodynamic quantities. 

In the context of the 1-D Ising model, information geometry has been studied by Ruppeiner \cite{rup1981} and later by Janyszek and Mrugala, \cite{jm}. 
We briefly recall their results. We start with the thermodynamic potential per 
unit spin in the ferromagnetic case (with unit coupling constant), which is given by
\begin{equation}
\Phi = -\frac{1}{\beta}{\rm ln}\left[e^{\beta}{\rm cosh}\left(\alpha\right) + \left(e^{2\beta}{\rm sinh}^2\left(\alpha\right) + e^{-2\beta}\right)^{1/2}\right]
\label{gibbs1dising}
\end{equation}
where $H$ and $T$ denote the applied magnetic field and the temperature respectively, and $\alpha = H/T$, $\beta = 1/T$ (the Boltzmann's constant has been set to be unity). 
From eq.(\ref{gibbs1dising}), the authors of \cite{jm} showed that in the $\left(\alpha,\beta\right)$ coordinates, the scalar curvature of the thermodynamic 
state space is given by the simple analytical expression
\begin{equation}
|R| = \frac{{\rm cosh}\left(\alpha\right)}{\left({\rm sinh}^2\left(\alpha\right) + e^{-4\beta}\right)^{1/2}} + 1
\label{curvising}
\end{equation}
This also agrees with the result of \cite{rup1981}. In the zero field limit, the curvature diverges at $T=0$ indicating the sole critical point of this theory. Also, it can be checked that
\begin{equation}
\xi|_{H=0} = \frac{1}{|{\rm ln}\left({\rm tanh}\left(\beta\right)\right)|} \simeq \frac{1}{2}|R|_{H=0}
\end{equation}
In order to understand the Widom line, we study information geometry of the 1-D Ising model away from criticality for non-zero $T$ and $H$. 
We compute the maxima of the correlation length (equivalently, $|R|$) in the supercritical region, and contrast it with the maxima of other response functions. 
We first record the standard expression for the magnetization per spin
\begin{equation}
m = \frac{{\rm sinh}\left(\alpha\right)}{\left({\rm sinh}^2\left(\alpha\right) + e^{-4\beta}\right)}
\end{equation}
The specific heat at constant magnetization is then obtained by differentiating the entropy obtained from eq.(\ref{gibbs1dising}), while keeping $m$ fixed, 
\begin{equation}
 C_m = \frac{4e^{2\alpha}}{T^2\left(e^{4\beta}{\rm sinh}^2\left(\alpha\right) + 1\right)^{1/2}\left(e^{2\beta}{\rm cosh}\left(\alpha\right) + 
 \left(e^{4\beta}{\rm sinh}^2\left(\alpha\right) + 1\right)^{1/2}\right)^2{\rm cosh}\left(\alpha\right)}
 \end{equation}
with a lengthy expression for the specific heat $C_H$ at constant applied field $H$ which is not reproduced here. These
of course reduce to the standard expression $C_{m,H}|_{H=0} = \beta^2{\rm sech}^2\beta$
in the zero field case. 
\begin{figure}[t!]
\centering
\includegraphics[width=2.8in,height=2.3in]{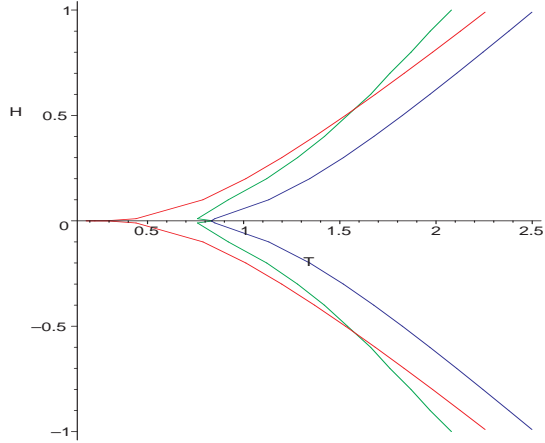}
\caption{``Widom lines'' for the 1-D Ising model on the $H-T$ plane. The red curve is the locus of maxima of the scalar curvature $R$. The blue and the green curves are
the maxima of the specific heats $C_M$ and $C_H$ respectively.}
\label{1dising}
\end{figure}
Now, we calculate the maxima of $|R|$ as a function of temperature. From eq.(\ref{curvising}), it can be seen that such maxima occur for a given value of $H$ 
when the the temperature $T$ satisfies the equation
\begin{equation}
{\rm tanh}\left(\frac{H}{T}\right) = \frac{2}{H\left(e^{4/T} - 1\right)}
\end{equation}
and a similar analysis can be carried out to determine the maxima of the specific heats $C_m$ and $C_H$. \footnote{We will loosely refer to the locus of 
maxima of the correlation length and the other response coefficients as the ``Widom lines.''} 
Numerical solutions of these have been plotted
in fig.(\ref{1dising}), where the red, green and blue curves denote the locus of maxima of $|R|$, $C_H$, and $C_m$ respectively.  We observe that near the critical 
point, i.e for very small values of the applied magnetic field $H$, the maxima of the specific heat do not asymptote to $T=0$, but reaches a limiting temperature 
$T\simeq 0.8$. The maxima of $|R|$ (i.e the correlation length $\xi$) for such small values of $H$ also does not asymptote to zero, but goes to a limit 
$T \simeq 0.18$ for $H = 10^{-5}$. It is interesting to note that the locus of maxima of the correlation length does not begin from the critical point for the 1-D Ising
model. As we will see, this feature will not be present in any of the mean-field models that we will analyse. 

Before concluding this section, let us summarize the main results here.
We have considered the 1-D Ising model from the point of view of information geometry. Analysing of the same away from the critical point, we have located the
locus of maxima of the correlation length $\xi$ and the specific heats $C_m$ and $C_H$, in the $T-H$ plane. 
Our analysis points to the fact that in this simple example, the Widom lines associated with the locus of maxima of $R$ and also the specific heats do not 
seem to converge to the critical point ($T$= 0) in the limit $H \to 0$.

\section{Information Geometry and The Curie-Weiss Mean Field Model}

In this section, we study the information geometry of the classical mean-field Curie-Weiss (CW) ferromagnetic model. This model has been studied extensively in the past,
and details can be found in standard textbooks \cite{thompson},\cite{dc}. We will adhere to the notations of \cite{thompson}. In the context of information geometry,
the CW model was first studied by Janyszek and Mrugala in \cite{jm}. These authors established the divergence of the scalar curvature close to the critical point. We seek to 
understand this model away from criticality, and study first order phase transitions, via the geometry of the thermodynamic state space. 

As is well known, in the thermodynamic limit, the free energy of the CW model is given by
\begin{equation}
{\mathcal G} = -T\left({\rm ln}2 + {\rm max}f\left(m\right)\right)
\end{equation}
where the Boltzmann constant has been set to unity, and $m$ is the magnetization per spin, which solves the equation 
\begin{equation}
m = {\rm tanh}\left(\frac{T_c}{T}m + \frac{H}{T}\right)
\label{soleta}
\end{equation}
Here, $T$ is the temperature, $T_c$ its critical value, and $H$ is the applied magnetic field. In order to write this in terms of the temperature 
and the thermodynamic extensive variable, $m$, we perform a Legendre transform and write the free energy as \cite{jm}
\begin{equation}
f = -T{\rm ln}2 - \frac{1}{2}T_cm^2 + \frac{T}{2}{\rm ln}\left(1 - m^2\right) + Tm{\rm tanh}^{-1}m
\label{cwhelm}
\end{equation}
with $\left(\frac{\partial f}{\partial m}\right)_T = H$. This can be used to calculate the scalar curvature of the equilibrium thermodynamic state space,
but note that the entropy per spin is given by eq.(\ref{cwhelm}) as
\begin{equation}
s = {\rm ln}2 -\frac{1}{2}{\rm ln}\left(1-m^2\right) - m{\rm tanh}^{-1}m
\label{cwentropy}
\end{equation}
As can be immediately seen from eq.(\ref{cwentropy}) and eq.(\ref{entropyrep}), 
information geometry becomes trivial in this context, since the entropy is a function of a single variable, and is not 
amenable to a Riemannian geometric analysis. In order to remedy the situation, the authors of \cite{jm} proposed to modify the Hamiltonian, which in its original
form for $N$ spins, is given by
\begin{equation}
{\mathcal H} = -\frac{T_c}{N}\sum_{i<j}s_is_j - H\sum_i s_i,~~~i,j=1\cdots N
\end{equation}
by an additional term that corresponds to the mechanical energy of the lattice. In particular, the modification seeks to retain a non zero specific heat for the model,
which is otherwise zero as can be seen from eq.(\ref{cwentropy}). Doing this modification {\it ad hoc} essentially amounts to assuming that
\begin{equation}
-T\left(\frac{\partial^2f}{\partial T^2}\right)_{m} = C_L\left(T\right)
\end{equation}
where $C_L$, the ``lattice specific heat'' (not to be confused with the specific heat at constant $m$) is an apriori unknown function of the temperature. As we show below, 
this modification nevertheless captures the correct information theoretic properties of the system. 

The line element is now given by \cite{jm}
\begin{equation}
dl^2 = \frac{C_L}{T^2}dT^2 + \frac{1}{T}\left(\frac{\partial^2f}{\partial m^2}\right)_Tdm^2
\label{linecw}
\end{equation}
where we have used the free energy representation, with coordinates $\left(T,m\right)$ \cite{rupp}. \footnote{Note that the second term in eq.(\ref{linecw}) is also related to
the inverse of the magnetic susceptibility, i.e $\left(\frac{\partial^2f}{\partial m^2}\right) = \frac{1}{\chi_T} = \frac{1-T-m^2}{1-m^2}$}
A simple calculation then shows that the scalar curvature for this model is given by $R = {\mathcal A}.{\mathcal B}$, where
\begin{eqnarray}
{\mathcal A} &=& \frac{1}{2}\frac{T_c\left(1-m^2\right)}{C_L^2\left(T - T_c\left(1-m^2\right)\right)^2} \nonumber\\
{\mathcal B} &=& T_cm^2\left(TC_L' + C_L\right) - T_c\left(C_L + TC_L'\right) + T\left(2C_L + TC_L'\right)
\label{curvforcw}
\end{eqnarray}
\begin{figure}[t!]
\begin{minipage}[b]{0.5\linewidth}
\includegraphics[width=2.8in,height=2.3in]{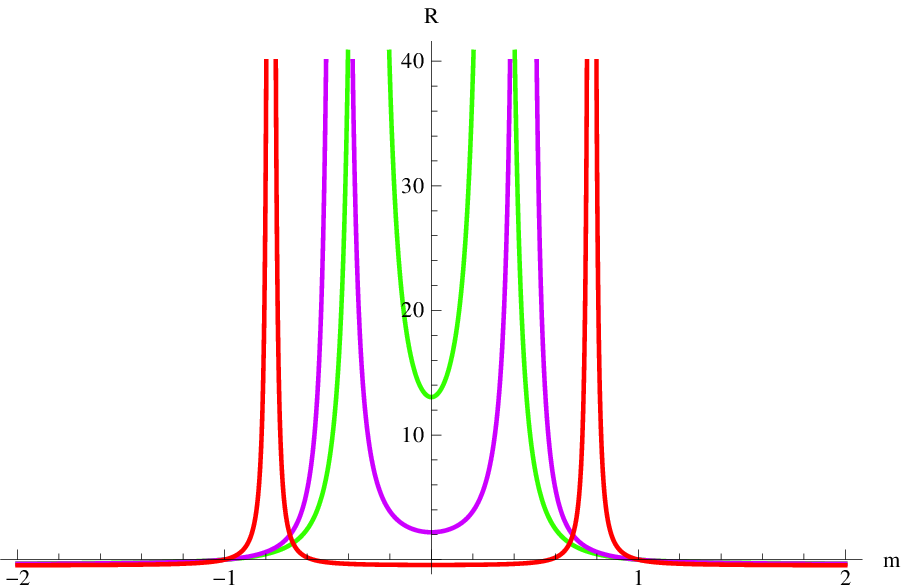}
\caption{Isothermal $R$ vs $m$ for the Curie-Weiss model, below criticality. The red, magenta and green curves correspond to 
$T=0.4$, $0.8$ and $0.9$ respectively. The critical temperature has been chosen to be $T_c=1$.}
\label{cw1}
\end{minipage}
\hspace{0.2cm}
\begin{minipage}[b]{0.5\linewidth}
\includegraphics[width=2.8in,height=2.3in]{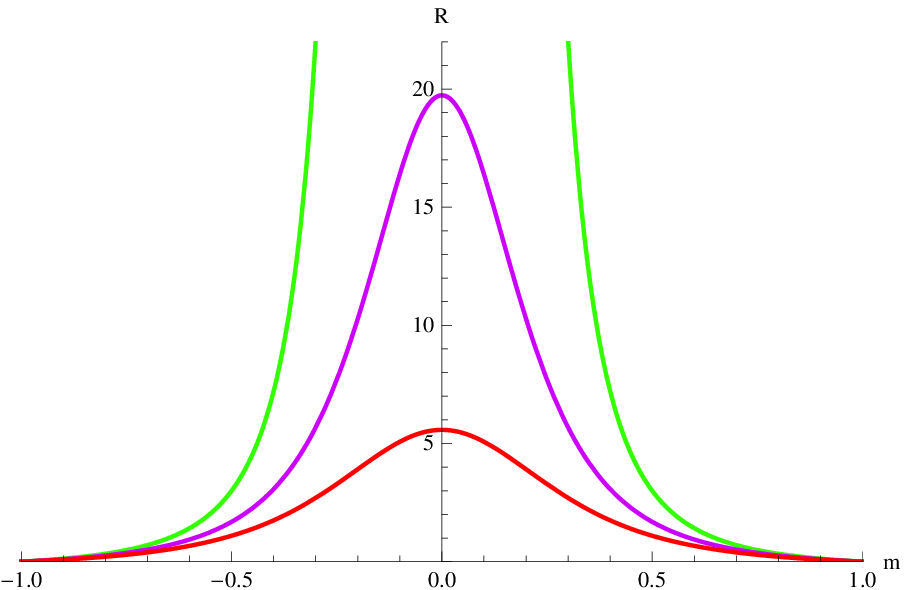}
\caption{Isothermal $R$ vs $m$ for the Curie-Weiss model, beyond criticality. The green line shows a diverging curvature at the origin
for $T=T_c=1$. The magenta and red curves correspond to $T=1.1$ and $1.2$ respectively.}
\label{cw2}
\end{minipage}
\end{figure}
Here, the primes denote derivatives of $C_L$ with respect to the temperature $T$.
In particular, we see that the curvature diverges as $T \to T_c$ (i.e $m \to 0$), as derived in a slightly different way in \cite{jm}. 
We now present a general analysis of $R$, valid for non-zero $T$ and $H$, away from criticality.  
The form of the curvature scalar in eq.(\ref{curvforcw}) is useful when we assume a specific form of $C_L$ and a value of $T_c$. For simplicity, we assume $T_c=1$ and 
a power law, $C_L\left(T\right) = 1 + T + T^2$. This is only for illustration, and as can be checked, any other value of the critical temperature or any other 
regular functional form of $C_L$ will not alter our discussion below. 
With these assumptions, the curvature simplifies to 
\begin{equation}
R = \frac{1}{2}\frac{\left(1-m^2\right)\left[m^2\left(1 + 2T + 3T^2 \right) + 4T^3 - 1\right]}{\left(1-m^2-T\right)^2\left(1+T + T^2\right)^2}
\label{RCW}
\end{equation}
We also calculate the specific heat at constant applied field $H$, using 
\begin{equation}
H = T{\rm tanh}^{-1}m - T_cm
\label{trans}
\end{equation}
and obtain from eq.(\ref{cwentropy}), for $T_c=1$,
\begin{equation}
C_H = T\left(\frac{\partial s}{\partial T}\right)_H =\frac{ T\left({\rm tanh}^{-1}m\right)^2\left(m^2 - 1\right)}{1-m^2 - T}
\label{chcw}
\end{equation}
We observe that the curvature scalar blows up wherever $C_H$ does, as expected, for $T < T_c$. (However, $C_H$ does not diverge at the critical point).
Further, the denominator of $R$ behaves as the product of 
the square of $C_L$ and the square of the denominator of $C_H$. 
We keep this in mind, as this seems to be a universal feature in mean-field theories.

Now, we study the behavior of $R$ as a function of $m$. This is shown in fig.(\ref{cw1}) for temperatures below criticality
and in fig.(\ref{cw2}) for temperatures at and beyond criticality. For $T<T_c$, we find that $R$ diverges symmetrically on the $m$-axis for a given value of 
temperature. This value of $m$ can, in fact, be identified with the turning point of the magnetic isotherms in the $H-m$ plane. We have not shown these isotherms
here, but we note that for values of $m$ that lies between the two divergences, the specific heat $C_H$ is negative, i.e, the system
becoms unstable. The two divergence on the $m$-axis get closer as we approach $T_c = 1$, at which point they merge into a single divergence at $m=0$. 
Beyond criticality, the situation is depicted in fig.(\ref{cw2}). Here we see that isothermal $R$ has a maximum at $m=0$, with its value at the maximum decreasing as we move 
away from criticality.
\begin{figure}[t!]
\begin{minipage}[b]{0.5\linewidth}
\includegraphics[width=2.8in,height=2.3in]{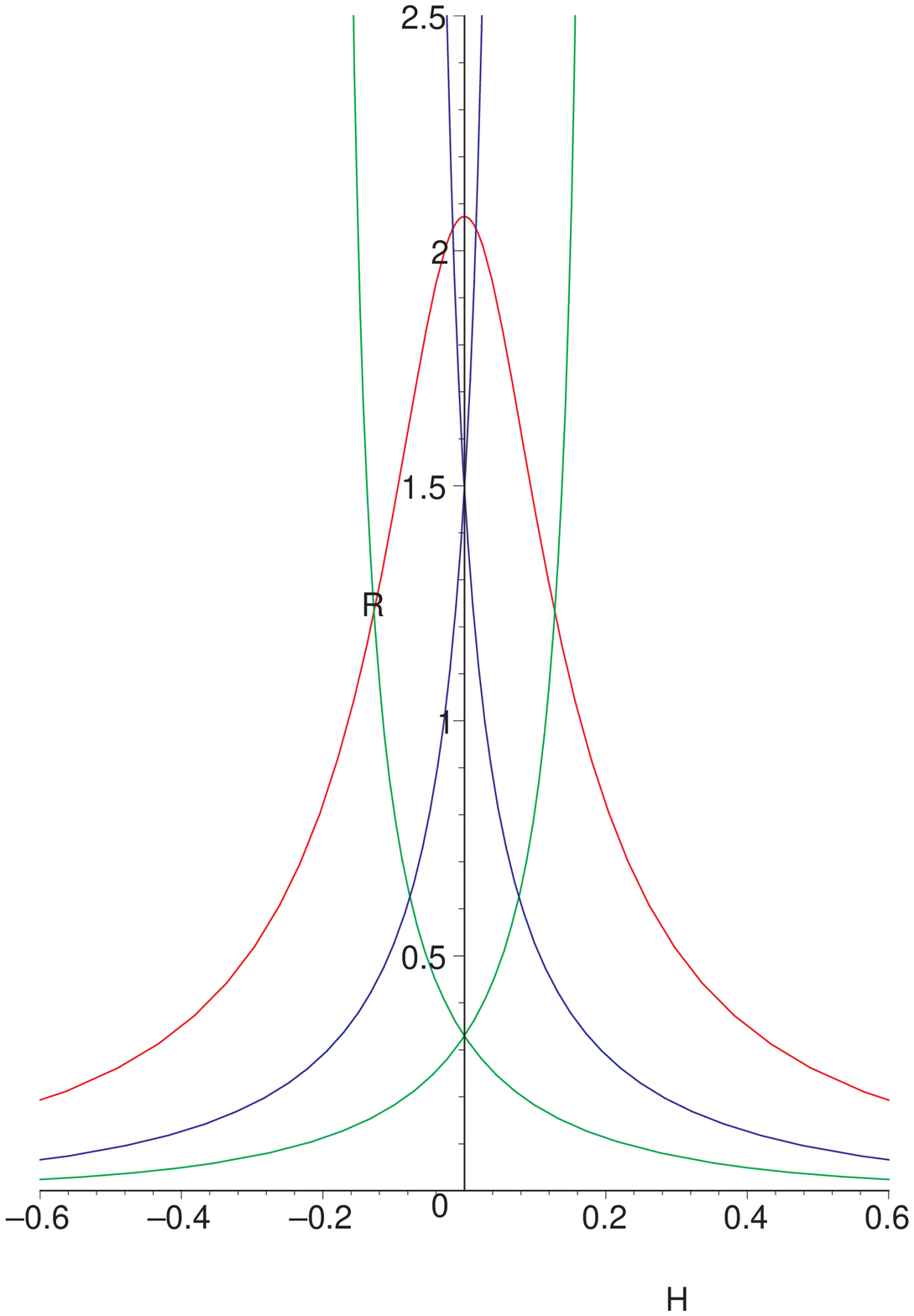}
\caption{$R$ vs $H$ for isotherms the Curie-Weiss model. The green curves correspond to $|R|$ in the two physical regions at $T=0.6$, and the blue curves are 
for $T=0.8$. The red curve corresponds to the isotherm $T = 1.35$. Crossing of the physical branches of $|R|$ indicate phase transition, always at $H=0$.}
\label{cwrcrossing}
\end{minipage}
\hspace{0.2cm}
\begin{minipage}[b]{0.5\linewidth}
\includegraphics[width=2.8in,height=2.3in]{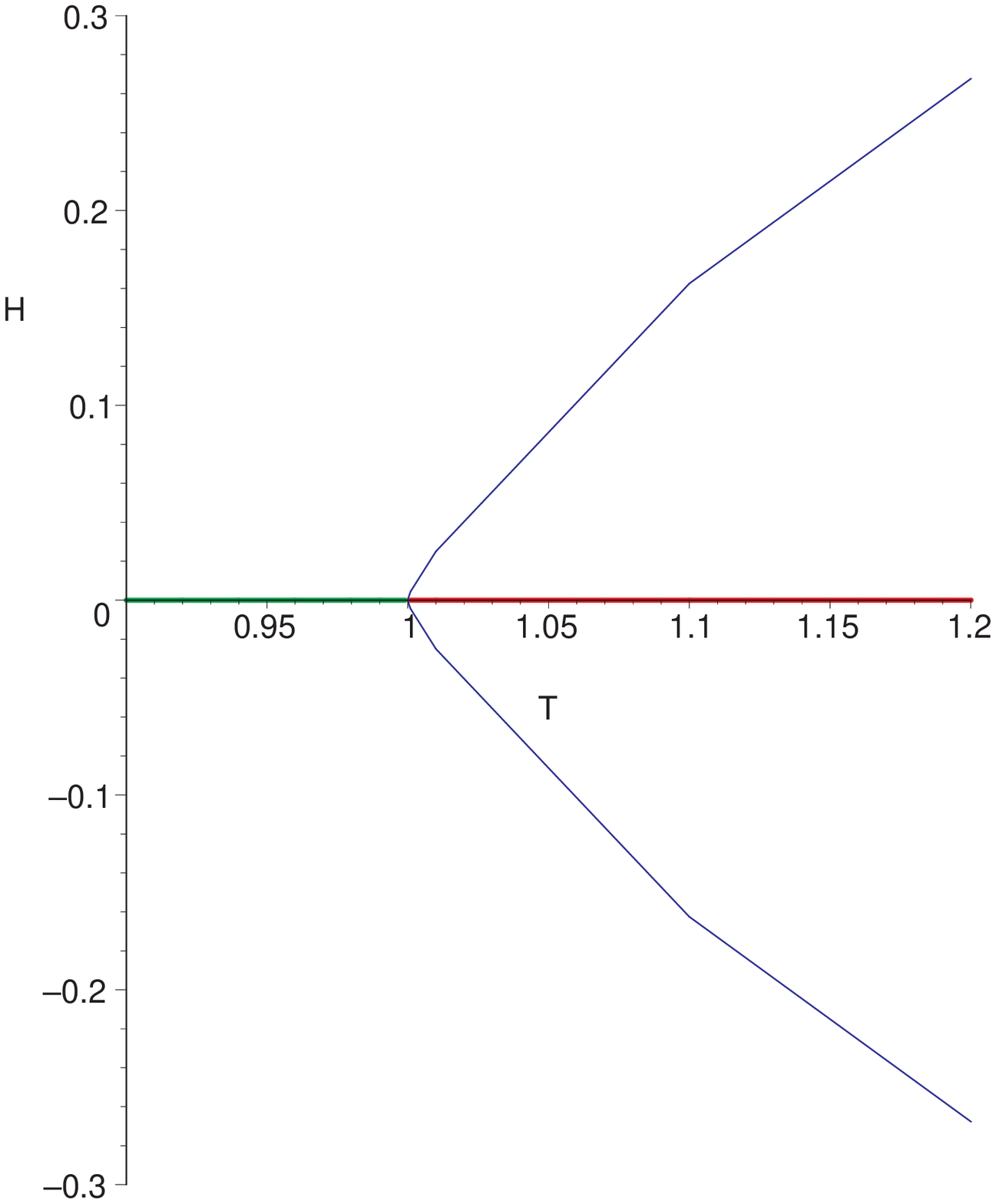}
\caption{Loci of maxima for the CW model. The green line along the $H$-axis denotes phase co-existence and terminates at $T_c=1$. The red line, along $H=0$ are the
maxima of $|R|$ (and $\chi_T$), the Widom line for various values of $T$. The two blue lines are the symmetric maxima of $C_H$ as a function of $T$.}
\label{cwwidom}
\end{minipage}
\end{figure}

It is more useful to consider the behavior of $R$ as a function of the applied magnetic field, $H$, with $m$ being treated as a parameter.  
We can guess the result by exploiting the symmetry of the situation. From the expression for $R$ of eq.(\ref{RCW}), we see that $R\left(m\right) = R\left(-m\right)$.
For two physical branches (where $C_H$ is positive), the $R$'s should thus ``cross'' where $m=0$. From eq.(\ref{trans}), this occurs when $H=0$. 

We show this in fig.(\ref{cwrcrossing}), for the isotherms $T=0.6$ (green), $T=0.8$ (blue) and $T=1.35$ (red), which we now explain this in details. 
Take, for example, the isotherm $T=0.8$ of fig.(\ref{cw1}) (the magenta curve). For this isotherm, the two blue curves of fig.(\ref{cwrcrossing}) are plots of $R$ in the 
the regions where $C_H$ is positive, with the curve that asymptotes to infinity for positive values of $H$ are for negative values of $m$ and the
other one is for corresponding positive values of $m$. Specifically, from fig.(\ref{cw1}), the $T=0.8$ $R$-isotherm diverges at $m=\pm 0.4472$
(as can be seen from eq.(\ref{RCW}). The blue curve
in fig.(\ref{cwrcrossing}) that diverges at a positive value of $H$ is a parametric plot for $m < -0.4472$ and the one that diverges at a negative value of $H$
is for $m > 0.4472$.  From eq.(\ref{trans}), the divergence occurs at $H = \pm 0.0622$. This is exactly as shown in fig.(\ref{cw1}). Beyond $T_c$, there is no
$R$-crossing. 

We thus see that below the critical temperature, $R$ has two physical branches, which cross at $H=0$. Indeed, as alluded to in the
introduction, this implies equality of correlation lengths at $H=0$ which we interpret as the phase transition, with the residual magnetization being the values of $m$
where $R$ diverges for this temperature. Of course beyond $T_c$, there is no residual magnetization, as can be seen from the behavior of $R$ in fig.(\ref{cw2}).
This is the information geometric description of discontinuous transitions in magnetic systems, and agrees with standard results. 

As we move closer to $T_c$, the value of $R$ at 
which the physical branches cross, increases. This can be seen by comparing the green and blue curves in fig.(\ref{cwrcrossing}), which denote physical branches of $R$ 
at $T=0.6$ and $T=0.8$ respectively.  
At $T_c = 1$, the crossing point is pushed to infinity. Beyond $T_c$, $R$ shows a maxima, which is always at $H=0$, which can again be
justified using symmetry arguments. The location of the Widom line is hence 
along the $H$-axis, evan away from criticality. This is depicted in fig.(\ref{cwwidom}), where the green line ($H$=0), is the first order line that 
culminates at $T_c=1$, and the locus of maxima of $R$ is its continuation, along the red line. 

To contrast the behavior of the maxima of $R$ with other response coefficients, we have also plotted, in fig.(\ref{cwwidom}), the locus of maxima of the specific
heat $C_H$. We find that unlike $R$ (and the susceptibility $\chi_T$, whose locus of maxima is the same as that of $R$), 
$C_H$ shows two symmetric maxima for a given temperature on the $H$-axis. All maxima converge to the critical point for very small values of $H$. 

To conclude this section, we summarize the main results. Here, we have provided a novel characterisation of first order processes in a simple mean-field magnetic model,
the CW model of ferromagnetism, via information geometry. We have seen that equality of the correlation length $\xi$ of co-existing phases accurately predicts known 
behavior of this model, near or away from criticality. The interpretation of the scalar curvature of the thermodynamic state space $R$ further allows us to calculate the 
Widom line as the locus of maxima of the correlation length $\xi$ (via $R \sim \xi^d$) which is shown to lie on the $H$-axis, as an expected continuation of the phase co-existence 
line, in the $T-H$ plane. Such loci of maxima for the specific heat $C_H$ however do not lie on the $H$-axis. 
We further note that our information geometric study of first order processes in the CW model has essentially similar features as mean-field liquid-gas 
systems \cite{rsss}, as expected, and indicate the wide applicability of this method in studying phase transitions. As an illustration of this, we now proceed to study
liquid systems. 

\section{Information Geometry and Liquid Systems}

We now apply our information geometric method to a simplified toy example in the thermodynamics of 
liquid-liquid phase transitions. Although the literature on the topic is vast (see, for example, \cite{vss} for a recent analysis in silica) analytic insight in such systems is often 
difficult to provide in practice. We choose a simple model that nevertheless captures 
the essential physics. The model that we study has internal energy 
\begin{equation}
U = A\left(V\right) + B\left(V\right)T^{3/5} + kT
\label{internal}
\end{equation}
where $k$ is related to Boltzmann's constant. 
This form of the energy is motivated from the fact that the potential energy $U \sim T^{3/5}$ for some simple liquids \cite{rosen},\cite{ssp}. 
Here, $A\left(V\right)$ and $B\left(V\right)$ are functions of the volume $V$, which can be taken to be polynomial fits, determined from simulation data. The exact nature
of these functions for liquid silica has been dealt with extensively in \cite{ssp} (where a similar form of the internal energy was used), and they were fitted to fourth 
order polynomials in the volume. For analytical tractability, it is enough for us to choose a simpler situation, and we assume
\begin{eqnarray}
A\left(V\right) &=& A_0 + A_1V + A_2V^2\nonumber\\
B\left(V\right) &=& B_0 + B_1V
\label{liqcoeffs}
\end{eqnarray}
where $A_i,~i=0\cdots 2$ and $B_j,~j=0,1$ are constants. 
Inclusion of a quadritic term in $B\left(V\right)$ does not allow for analytic handling of the model, and we have dropped this for the time being. 
Indeed, this simple choice of the functions $A$ and $B$ will serve to illustrate the main features for any physical system with a $T^{3/5}$
dependence of the internal energy, and it can be checked that the inclusion of a quadritic term in $B\left(V\right)$ will not alter the qualitative aspects of our discussion. 
We will also assume throughout that the units are properly chosen, as in \cite{ssp}.
\begin{figure}[t!]
\begin{minipage}[b]{0.5\linewidth}
\includegraphics[width=2.8in,height=2.3in]{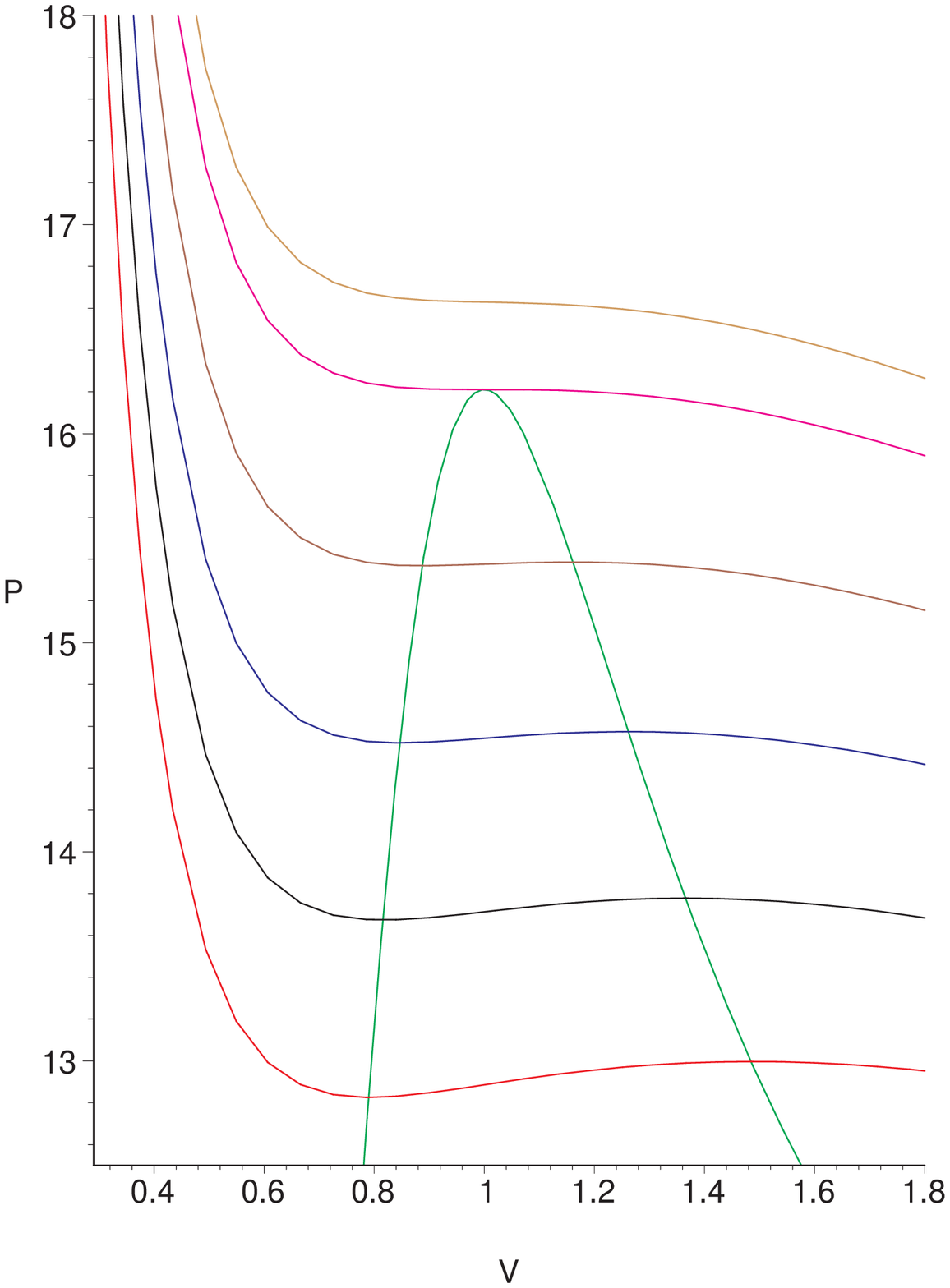}
\caption{Isotherms in the $V-P$ plane for various values of temperature.}
\label{liqisotherms}
\end{minipage}
\hspace{0.2cm}
\begin{minipage}[b]{0.5\linewidth}
\includegraphics[width=2.8in,height=2.3in]{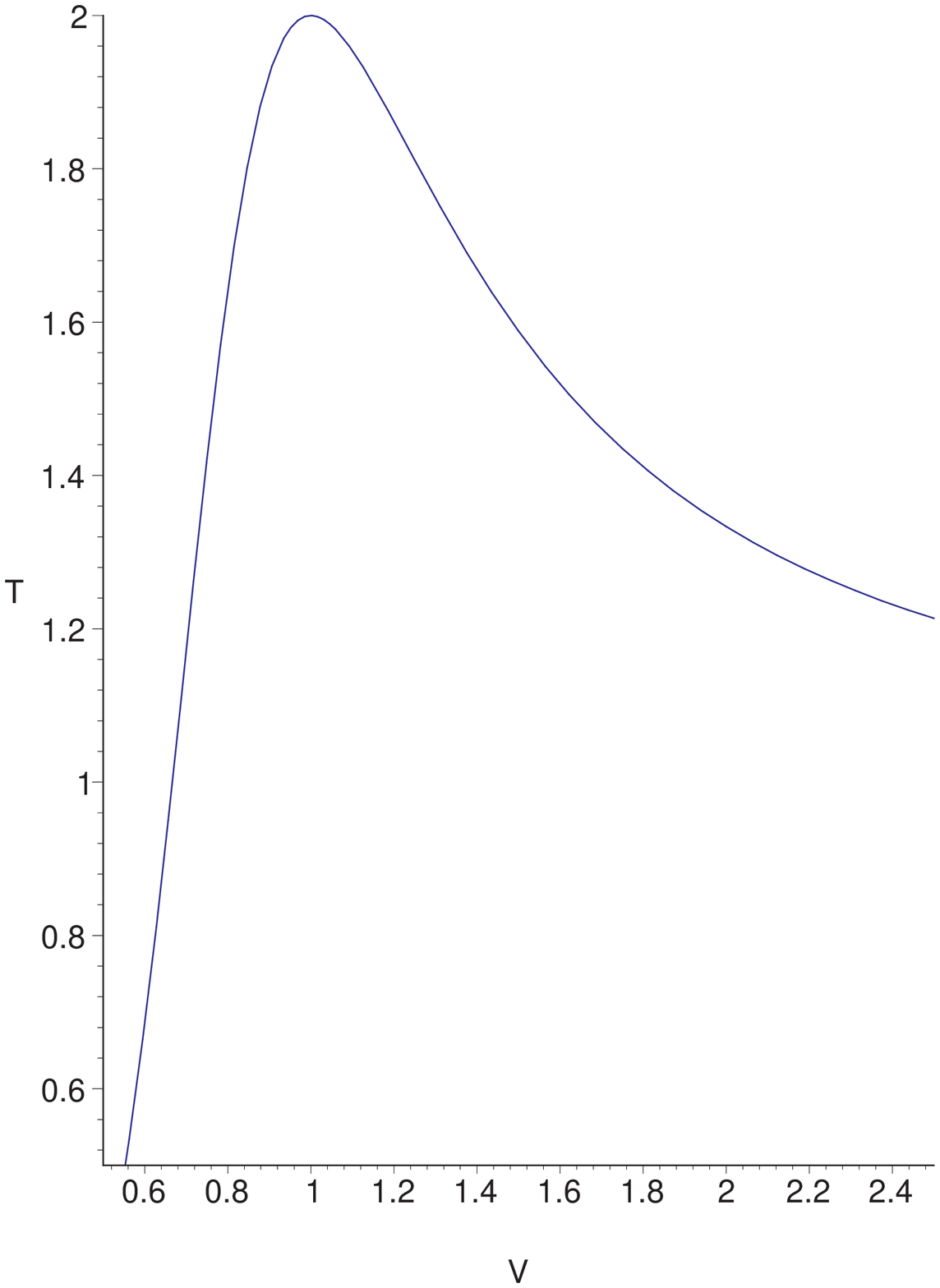}
\caption{Meta-stability region in the $T-V$ plane.}
\label{liqmetastable}
\end{minipage}
\end{figure}
Given the internal energy $U$, one can compute the entropy of the system by textbook methods \cite{callen}. In particular, starting from a reference entropy $S\left(T_0,V_0\right)$ at 
a reference temperature $T_0$, and a reference volume $V_0$, the entropy at arbitrary values of the temperature and volume 
can be calculated by adding its change along and isotherm and along an isochore. This requires the expression for the pressure for the reference temperature, which can
be again obtained as a polynomial fit in the density \cite{ssp}. In our example, we choose
\begin{equation}
P|_{T=T_0} = C_0 + \frac{C_1}{V} + \frac{C_2}{V^2}
\label{liqpres}
\end{equation}
Where the $C$s are coefficients that can be calculated from molecular dynamics simulations. 
The calculation for the entropy has been pedagogically explained in \cite{ssp}, and we obtain 
\begin{eqnarray}
S &=& \frac{1}{2T^{2/5}V}\left[T^{2/5}\left(2V{\rm ln}T + 5VB\left(V\right) + \right.\right.\nonumber\\
&+& \left. \left.  2VA\left(V\right) + 2\left(V + C_0 V^2 + C_1V{\rm ln}V - C_2\right)\right) -3VB\left(V\right)\right]
\label{entropycw}
\end{eqnarray}
where we have set the constant $k$ in eq.(\ref{internal}), and the reference temperature $T_0$ to unity, without loss of generality. The Helmholtz free energy for this model
can be calculated from $F = U - TS$, and taking its derivative yields the pressure 
\begin{equation}
P = TP_0 -\frac{5}{2}B_1\left(T^{3/5} - T\right) + \left(T-1\right)\left(A_1 + 2A_2V\right)
\label{pressurecw}
\end{equation}
The curve of metastability can be obtained by setting
\begin{equation}
\left(\frac{\partial P}{\partial V}\right)_T = 2A_2\left(T - 1\right) - \frac{T}{V^3}\left(C_1V + 2C_2\right) = 0
\label{meta}
\end{equation}
The critical volume is calculated by solving for $T$ from the above equation and then using $\left(\frac{\partial T_m}{\partial V}\right) =0$, where
$T_m$ is the solution for the temperature obtained from eq.(\ref{meta}). These finally yield the critical temperature and volume :
\begin{equation}
T_c = \frac{54A_2C_2^2}{54A_2C_2^2 - C_1^3},~~~V_c = -3\frac{C_2}{C_1}
\label{tcvcliq}
\end{equation}
The constant volume specific heat for this model can be calculated to be
\begin{equation}
C_V = \frac{5T^{2/5} + 3\left(B_0 + B_1V\right)}{5T^{2/5}}
\end{equation}
And the isothermal compressibility is given by
\begin{equation}
K_T = -\frac{1}{V}\left(\frac{\partial V}{\partial P}\right)_T = \frac{V^2}{T\left(C_1V + 2C_2\right) -2A_2V^3\left(T-1\right)}
\label{kt}
\end{equation}
The expression for $C_P$, the specific heat at constant pressure is easy to obtain by differentiating eq.(\ref{entropycw}) while keeping the pressure in
eq.(\ref{pressurecw}) fixed, but gives a lengthy expressions, which we shall not
present fully here, but we note here that it has the same denominator as eq.(\ref{kt}), as expected. 
\begin{figure}[t!]
\begin{minipage}[b]{0.5\linewidth}
\includegraphics[width=2.8in,height=2.2in]{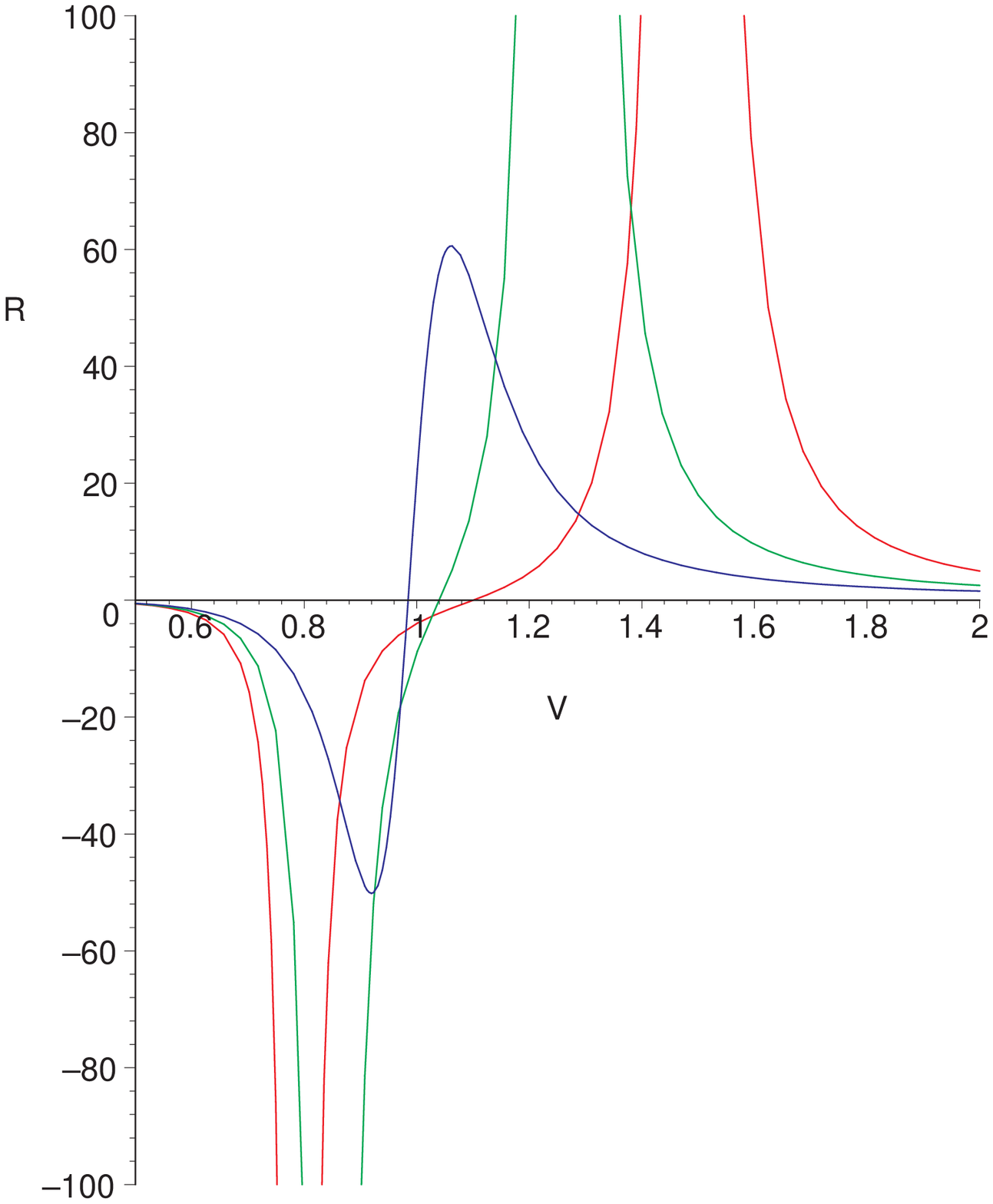}
\caption{$R$ as a function of volume for various isotherms. The red, green and blue curves correspond to $T=1.6$, $1.8$ and $2.1$ respectively. Beyond criticality,
$|R|$ shows two local maxima.}
\label{liqcurvature}
\end{minipage}
\hspace{0.2cm}
\begin{minipage}[b]{0.5\linewidth}
\includegraphics[width=2.8in,height=2.3in]{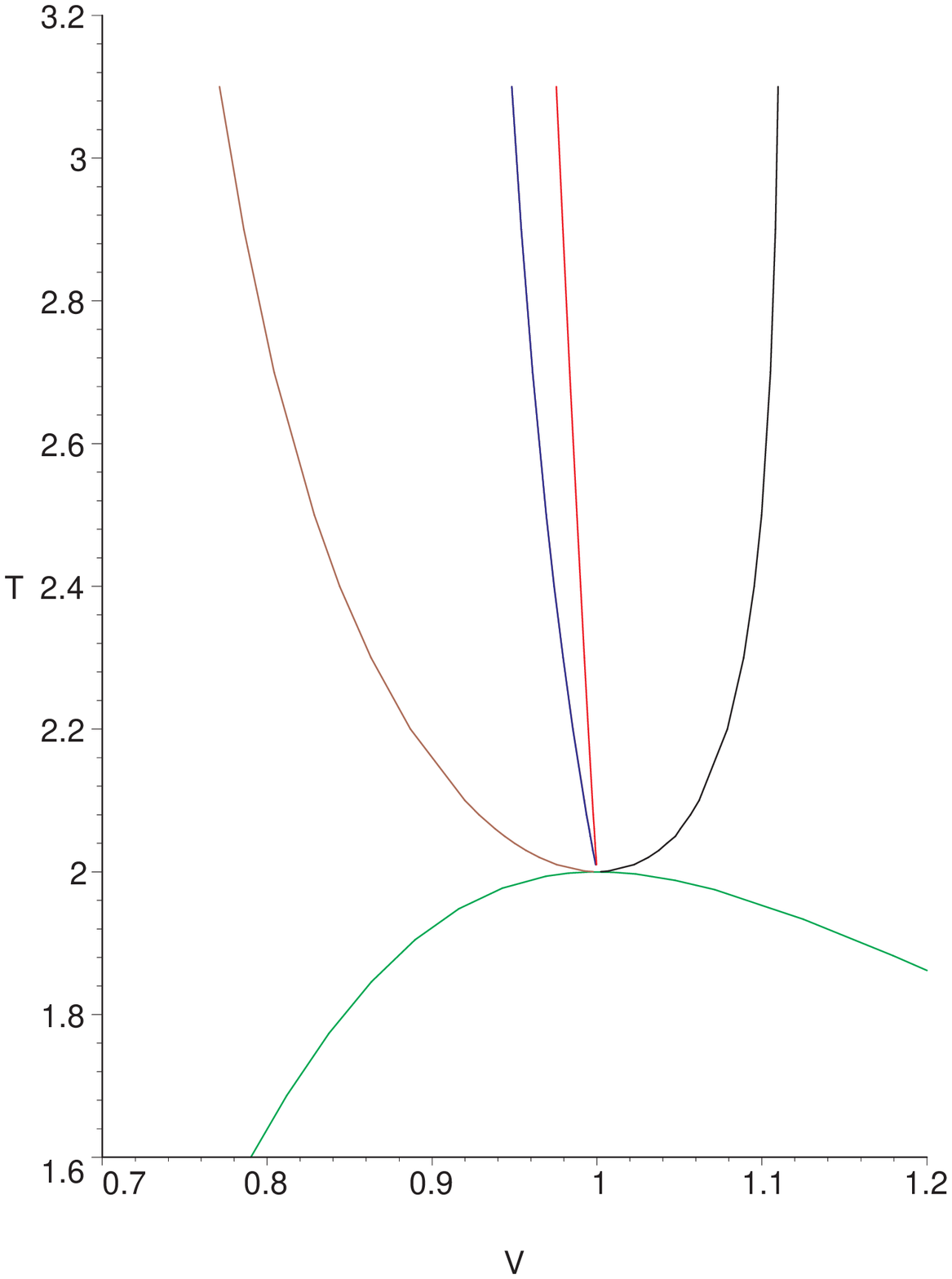}
\caption{Maxima of $|R|$ (black and brown curves), $C_P$ (red) and $K_T$ (blue) in the $V-T$ plane, beginning from the critical point in the spinodal
(green) curve in the $V-T$ plane.}
\label{liqmaxima}
\end{minipage}
\end{figure}

Our main interest here is in the scalar curvature of the equilibrium thermodynamic state space of this system. This can be calculated with the help of
either eq.(\ref{entropyrep}) or eq.(\ref{line}), and we obtain an expression for $R$ in terms of $T$ and $V$. Using eq.(\ref{line}) for example, we obtain a
diagonal form of the metric, given by the components
\begin{equation}
g_{TT} = \frac{C_V}{VT^2},~~~~~g_{\rho\rho} = \frac{V^2}{TK_T}
\end{equation}
where $\rho = 1/V$.
As required \cite{rupp}, these are certainly positive in the domain of interest, as is the quantity $\sqrt{g_{TT}g_{\rho\rho}}$. 
The expression for $R$ is too lengthy to reproduce here, but for the moment, let us note that its denominator is 
\begin{equation}
R_{\rm den}= \left[\left(5T^{2/5} + 3\left(B_0 + B_1V\right)\right)\left(T\left(C_1V + 2C_2\right) -2A_2V^3\left(T-1\right)\right)\right]^2
\end{equation}
implying that the curvature scalar diverges along the divergence of $K_T$ or $C_P$. Also note that the denominator of $R$ is equal to the product
of the square of the numerator of $C_V$ and the square of the denominator of $C_P$ (or $K_T$). This is a feature that we noticed in the Curie-Weiss model of
the previous section, and seems to be an universal property for mean-field systems. 

In order to make our results more tractable in general, we now make a choice of constants, and set $A_2=-1$, $B_0 = 5\times 10^3$, $C_0=10$, $C_1 = -3$,
with all the other constants set to unity. 
One can check that this ensures the positivity of physical quantities like temperature, volume, isothermal compressibility etc. in some domain, with our analysis
being valid in this domain.
With this choice of parameters, the metastability condition of eq.(\ref{meta}) yields, as a solution for the temperature,
\begin{equation}
T_m = \frac{2V^3}{2V^3 - 3V + 2}
\end{equation}
with the system being stable outside the region defined by the above equation, plotted in the $T-V$  plane. The critical temperature for our model is,
from eq.(\ref{tcvcliq}),  $T_c=2$, with 
$V_c=1$ and $P_c=16.21$ (in appropriate units). In fig.(\ref{liqisotherms}), we have shown some isotherms on the 
$P-V$ plane, along with the region of metastability for our model. The latter region is plotted in the $T-V$ plane in fig.(\ref{liqmetastable}). 
We now present our results on the scalar curvature graphically. We find that below $T_c$, isothermal $R$,
as a function of the volume, diverges at the boundaries of the metastability curve of fig.(\ref{liqmetastable}). This is shown in fig.(\ref{liqcurvature}). Consider, 
for example, the isotherm at $T=1.6$, plotted in red in fig.(\ref{liqcurvature}). For this temperature, $C_P$ diverges for $V=0.7899$ and $V=1.4844$. These
are also the values of $V$ for which $R$ diverges, and as can be seen from fig.(\ref{liqmetastable}) (or, equivalently, from eq.(\ref{meta}) after putting in the
constants), these are the values of $V$ between which the system becomes unstable. As we look at isotherms with higher temperature, for example the
$T=1.8$ isotherm plotted in green in fig.(\ref{liqmetastable}), the two divergences of $R$ come closer to each other, and at $T_c=2$, they merge at $V=1$. 
Beyond $T=1$, $|R|$ ($\sim \xi^3$) shows two maxima, as can be seen from the blue curve in fig.(\ref{liqmetastable}) which corresponds to an isotherm $T=2.1$. The Widom
line therefore has to be understood in more details here. 
\begin{figure}[t!]
\begin{minipage}[b]{0.5\linewidth}
\includegraphics[width=2.8in,height=2.2in]{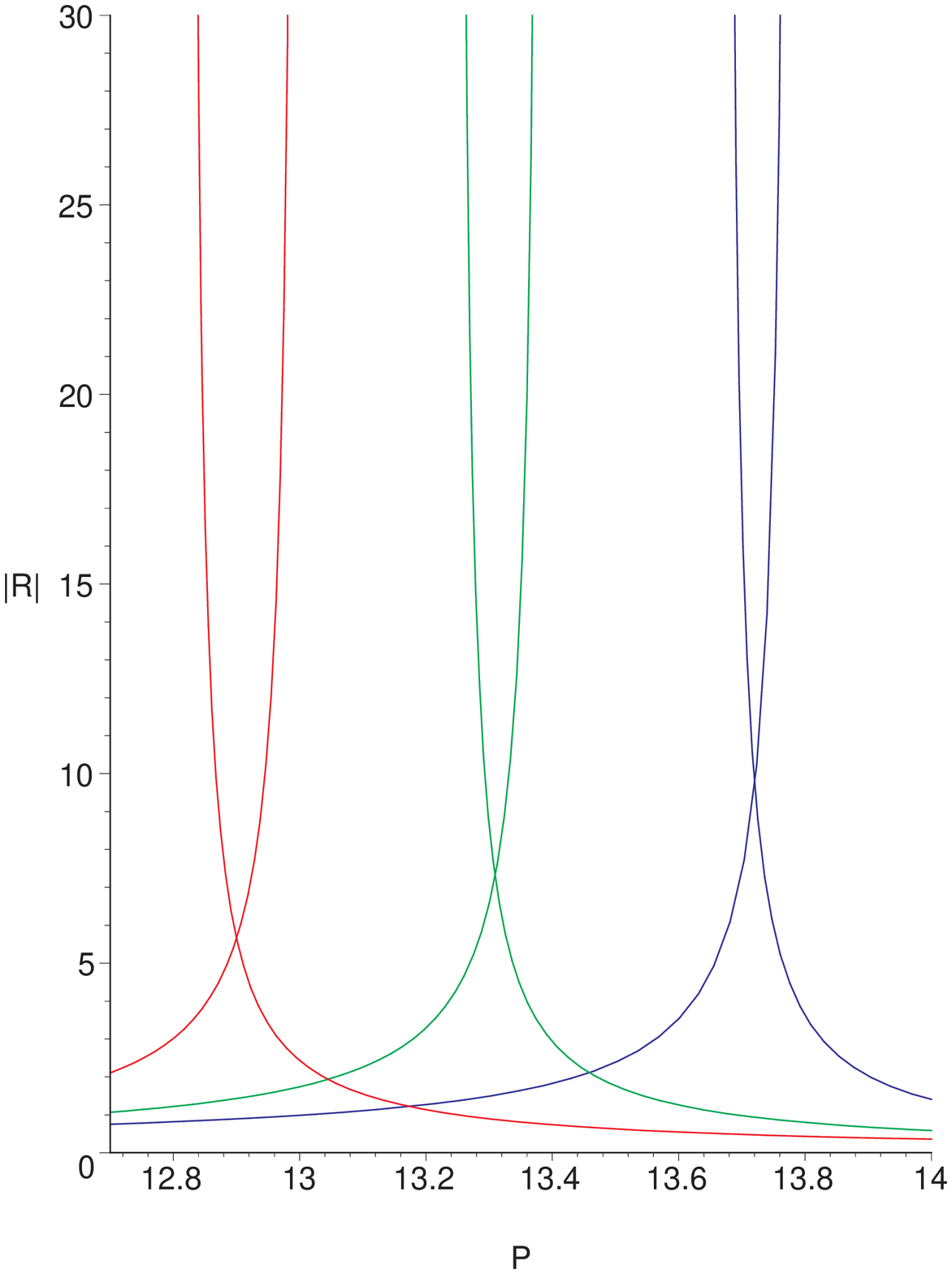}
\caption{Isothermal R-crossing as a function of $P$. Two curves of the same color denote $|R|$ in the physical regions for a given isotherm. 
The red curves correspond to $T=1.6$, the green to 
$T=1.65$ and the blue to $T=1.7$. The physical regions of the $P-V$ isotherms have been used for the plots.}
\label{Rcrossliq}
\end{minipage}
\hspace{0.2cm}
\begin{minipage}[b]{0.5\linewidth}
\includegraphics[width=2.8in,height=2.3in]{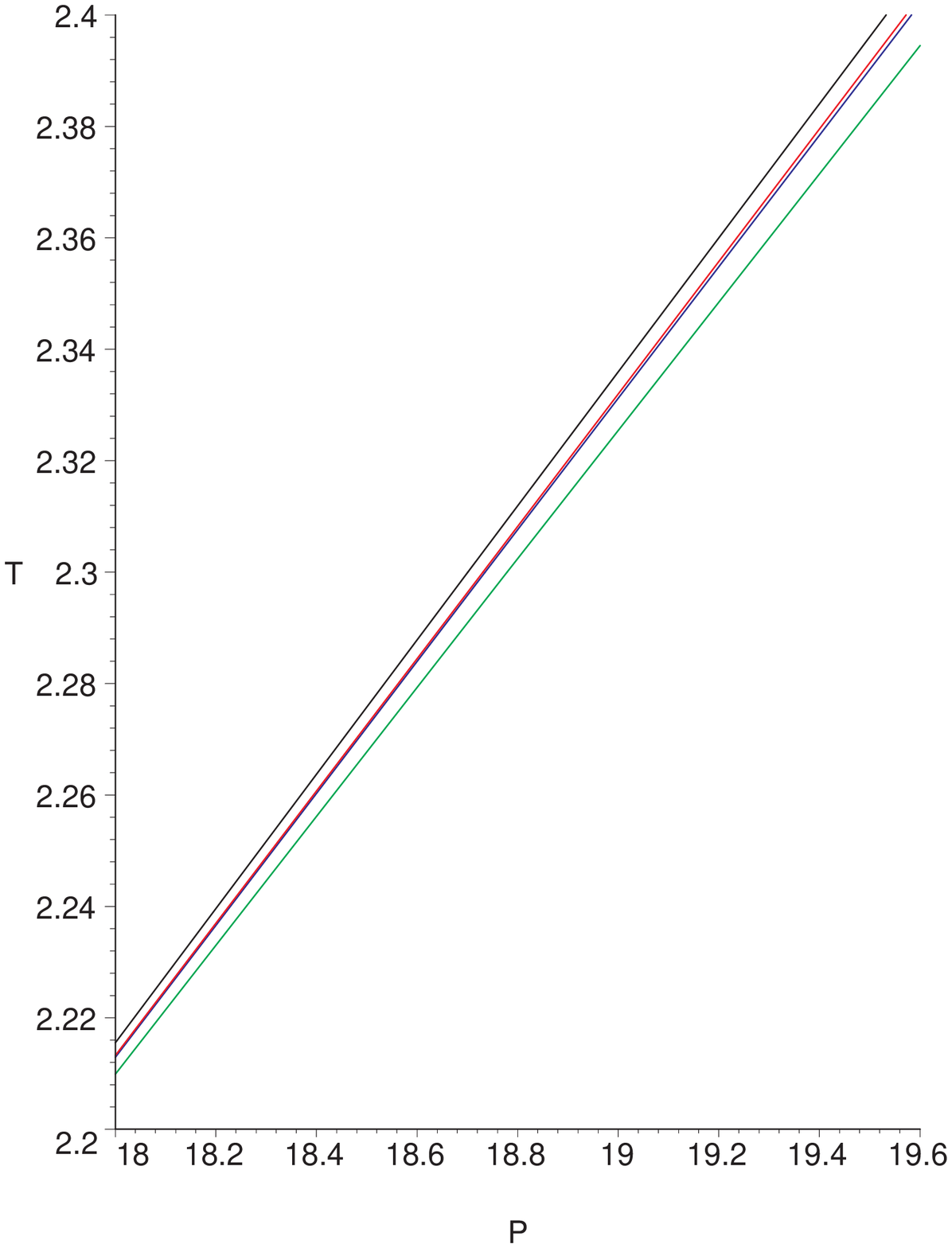}
\caption{Widom lines for the model liquid-liquid system. The green and black curves correspond to the two maxima of $|R|$. The red and blue curves
are almost indistinguishable and correspond to the maxima of $C_P$ and $K_T$ respectively. Near the critical pressure, these merge into a single line.}
\label{Widomliq}
\end{minipage}
\end{figure}

For this, beyond $T_c$, we study the maxima of $|R|$ by the equations
\begin{equation}
\left(\frac{\partial |R|}{\partial T}\right)_V =0,~~~\left(\frac{\partial |R|}{\partial V}\right)_P =0,~~~\left(\frac{\partial |R|}{\partial T}\right)_P =0
\label{Rmax}
\end{equation}
We find that whereas the maxima of $|R|$ obtained from the first of these relations (i.e isochoric maxima of $|R|$ with respect to the volume) is virtually indistinguishable from
the $K_T$ maxima line, those obtained from the second and third relations of eq.(\ref{Rmax}) indicate two local maxima for $|R|$ near criticality. The values of $V$ obtained
from both these relations (where $|R|$ maximises) are indistinguishable. Fig.(\ref{liqmaxima}) summarizes our results. Here, the red curve indicate the locus of
maxima of $C_P$, and the blue one is the corresponding quantity for $K_T$. The solution of the first relation of eq.(\ref{Rmax}) is identical to the blue curve. 
The solutions of the other two relations in eq.(\ref{Rmax}) give the brown and black curves. 

As in the case of the CW model of the previous section, instead of looking at isochores, 
it is more useful to look at the behavior of $|R|$ with respect to a thermodynamic intensive variable, such as the temperature or the pressure. 
In fig.(\ref{Rcrossliq}), we have shown the behavior of
isothermal $|R|$ as a function of the pressure, for various values of the temperature, with volume being the parameter in the plot. Consider, for example, 
the red curves in fig.(\ref{Rcrossliq}). These are isothermal plots of $|R|$ as a function of the pressure, for $T=1.6$, in the two physical domains of $V$ (as 
we have discussed) and correspond to the red curve in fig.(\ref{liqcurvature}). 
One of the red curves here (the one diverging at $P \sim 12.8$) corresponds to $V < 0.7899$ and the other red curve is 
for $V > 1.4844$. $|R|$ for the physical branches
is thus seen to cross at $P \sim 12.85$. This is thus the value of $P$ for which the correlation lengths of the coexisting phases become equal, and we  
interpret this as the pressure for which a first-order liquid-liquid phase transition occurs at $T=1.6$. Similar values of the pressure can be calculated using isotherms
for different values of $T$. The collection of all such points in the $T-P$ plane is the phase co-existence curve for the model. 

As we approach criticality, the crossing point of $R$ is pushed to infinity. Beyond criticality, there is no crossing, i.e a single phase 
exists, but $|R|$ shows two maxima with respect to the pressure, i.e there are two Widom lines that originate from the critical point. 
In fig.(\ref{Widomliq}), we have plotted the Widom lines for this system in the $P-T$ plane. Projected on this plane, the maxima of $|R|$ are shown in the green and black
curves and the other curves, which are the maxima of $C_P$ and $K_T$ become almost indistinguishable. All these curves converge at the critical point. 

Before we end, we summarize the main results in this section. Here, we have shown that information geometry can be effectively used to establish phase behavior
in liquid systems, in a simple way. The main inputs that went into our calculation is the power law behavior of the internal energy of eq.(\ref{internal}) and the coefficients
appearing in eqs.(\ref{liqcoeffs}) and (\ref{liqpres}). We stress here that although for illustrative purpose we have chosen a toy model, once these quantities are 
calculated in any liquid system, our method offers an easy way to determine phase behavior, and very importantly, the Widom lines, and may be used for
experimental predictions. We have also seen an important qualitative
difference between magnetic and liquid systems, namely that in the latter, there might be multiple isothermal Widom lines. We note here that this situation is 
different from liquid-gas systems as well, where there is a single locus of maxima of the correlation length $\xi$ for isotherms \cite{rsss}.

\section{Conclusions and Discussions}

In this paper, we have studied in details information geometry of magnetic and liquid systems. We first established the idea that equality of correlation lengths of 
co-existing phases, calculated via the scalar curvature of the equilibrium thermodynamic state space, is indicative of first order phase behavior in mean-field magnetic 
systems, where the curvature diverges appropriately at criticality. We then applied this to liquid systems and predicted liquid-liquid phase transition in a simple model. 
Our main conclusion here is that geometric techniques provide a universal new method of characterizing first and second order phase transitions, and can also be 
used to predict the behavior of the system beyond criticality, via the 
Widom line. Our results show that the definition of the latter as the locus of maxima of the correlation length is somewhat ambiguous in liquid systems, and can lead to
multiple lines, which originate from criticality. This is due to the $T^{3/2}$ behavior of the internal energy. This is to be contrasted with liquid-gas systems where the prediction of 
isobaric or isothermal Widom lines are unique \cite{rsss}. (For magnetic systems, such multiple locus of maxima occur instead for the specific heats).
For the 1-D Ising model, we have seen that the Widom line does not originate from the critical point. This however, can be attributed to the somewhat 
unphysical nature of the model, and this feature is not seen for any of the mean-field theories analysed in this paper.
It is an important question whether the Widom line is unique,
and future experiments will probably indicate which of the multiple lines is chosen by the system to distinguish between phases, beyond criticality. 

In liquid-gas systems, prediction of first order co-existence via the equality of the correlation length originates from the idea of Widom \cite{widom} that
near a first order transition, density fluctuations in one phase of a fluid results in the formation of a second phase. The thickness of the interface between the
two phases is then interpreted as the correlation length, which therefore should be equal measured in either phase \cite{rsss}. Our results indicate that a similar
phenomenon happens in magnetic and liquid systems as well, and this should be studied further. 

Compared to the more sophisticated computer simulation techniques to study liquid-liquid phase transitions advocated over the last two decades, ours is a simple method
which can be applied to any theoretical equation of state, or equivalently, experimental data.  Admittedly, we have chosen a simple scenario to illustrate our method, but
this can be easily generalised to more realistic situations, and the qualitative details should remain unchanged.  

It will be of interest to apply our technique to phenomenological models of liquid-liquid phase transitions. We note here that the predictions using the scalar curvature
remain valid as long as the numerical value of $|R|$ is more than typical molecular volumes. For liquid-gas systems, these restrict the use of geometric methods
beyond a certain range of temperature and pressure, though these are typically away from the scaling region. A similar analysis in liquid systems would 
involve  analysing data from molecular dynamics simulations. We leave this for a future publication.

\vspace{0.3in}
\begin{center}
{\bf Acknowledgements}
\end{center}
We wish to sincerely thank Amit Dutta, V. Subrahmanyam and K. P. Rajeev for extremely useful discussions.


\begin{thebibliography}{99}
\bibitem{callen}
H. B. Callen, {\it Thermodynamics and an Introduction to Thermostatistics}, John Wiley \& Sons, New York, 1985.
\bibitem{brodyhook}
D. C. Brody, D. W. Hook, ``Information geometry in vapor-liquid equilibrium,''
J. Phys. {\bf A42} (2008) 023001, {\tt arXiv : 0809.1166 [cond-mat.stat-mech]}.
\bibitem{weinhold}
F. Weinhold, ``Metric geometry of equilibrium thermodynamics,'' J. Chem. Phys. {\bf 63} (1975) 2479.
\bibitem{rupp}
G. Ruppeiner, ``Riemannian geometry in thermodynamic fluctuation theory,'' {\it Rev. Mod. Phys.} {\bf 67}, 605 (1995), erratum {\it ibid} {\bf 68}, 313 (1996). 
\bibitem{zanardi}
P. Zanardi, P. Giorda, M. Cozzini, ``Information-theoretic differential geometry of quantum phase transitions,'' Phys. Rev. Lett. {\bf 99} (2007) 100603.
\bibitem{rsss}
G.~Ruppeiner, A.~Sahay, T.~Sarkar, G.~Sengupta, ``Thermodynamic geometry, phase transitions, and the Widom Line,''
 {\tt arXiv:1106.2270 [cond-mat.stat-mech]}.
\bibitem{nist}
NIST Chemistry WebBook, available at the Web Site \\ http://webbook.nist.gov/chemistry/. 
\bibitem{stanley}
P. F. McMillan, H. E. Stanley, ``Fluid phases: going supercritical, Nature Physics {\bf 6}, (2010) 479.
\bibitem{stanley2}
L. Xu, P. Kumar, S. Buldyrev, S. Chen, P. Poole, F. Sciortino, H. E. Stanley, ``Relation between the Widom line and the dynamic crossover
in systems with a liquid-liquid phase transition,'' PNAS {\bf 102} (2005) 16558.
\bibitem{simeoni}
G. G. Simeoni, T. Bryk, F. A. Gorelli, M. Krisch, G. Ruocco, M. Santoro, T. Scopigno,
``The Widom line as the crossover between liquid-like and gas-like behavior in supercritical fluids,'' Nature Physics {\bf 6} (2010) 503.
\bibitem{widom}
Widom, B. The critical point and scaling theory. {\it Physica} {\bf 73}, 107 (1974).
\bibitem{rup1981}
G. Ruppeiner, ``Applications of Riemannian geometry to the thermodynamics of a simple fluctuating magnetic system,'' Phys. Rev. {\bf A 24} (1981) 488.
\bibitem{jm}
H. Janyszek, R. Mrugala, ``Riemannian geometry and the thermodynamics of model magnetic systems,'' Phys. Rev. {\bf A 39} (1989) 6515.
\bibitem{thompson}
C. J. Thompson, {\it Mathematical statistical mechanics}, Princeton University Press, Princeton NJ, 1972.
\bibitem{dc}
D. Chowdhury, D. Stauffer, {\it Principles of equilibrium statistical mechanics}, Wiley-VCH, 2000.
\bibitem{vss}
V. V. Vasisht, S. Saw, S. Sastry, ``Liquid-liquid critical point in supercooled silicon,'' Nature Physics {\bf 7} (2011) 549.
\bibitem{rosen}
Y. Rosenfeld, P. Tarazona, ``Density functional theory and the asymptotic high density expansion of the free energy of
classical solics and fluids,'' Mol. Phys. {\bf 95}, (1998) 141.
\bibitem{ssp}
I. Saika-Voivod, F. Sciortino, P. H. Poole, ``Computer simulation of liquid silica : Equation of state and liquid-liquid phase transition,''
Phys. Rev. {\bf E63}, 011202.
\end{thebibliography}
\end{document}